\documentclass[aps,twocolumn,groupedaddress,prl]{revtex4}
\usepackage{graphicx} 
\usepackage{amssymb,amsmath}

\begin{document}


\title{Realistic, quantitative descriptions of electron-transfer reactions:
\\diabatic free-energy surfaces from first-principles molecular dynamics}
\author{P. H.-L. Sit$^{1}$, Matteo Cococcioni$^{2}$ and Nicola Marzari$^{2}$}
\affiliation{$^{1}$Department of Physics,
Massachusetts Institute  of Technology, Cambridge, MA 02139, USA
\\
$^{2}$Department of Materials Science and Engineering,
Massachusetts Institute  of Technology, Cambridge, MA 02139, USA }

\date{\today}

\begin{abstract}
A general approach to calculate the diabatic surfaces for
electron-transfer reactions is presented, based on first-principles molecular
dynamics of the active centers and their surrounding medium. 
The excitation energy corresponding to the transfer of an electron
at any given ionic configuration (the Marcus energy gap) is accurately assessed 
within ground-state density-functional
theory via a novel penalty functional
for oxidation-reduction reactions 
that appropriately acts on the electronic degrees of freedom alone.
The self-interaction error intrinsic to common exchange-correlation
functionals is also corrected by the same penalty functional.
The diabatic free-energy surfaces
are then constructed from umbrella sampling on large ensembles of configurations.
As a paradigmatic case study, the self-exchange reaction between ferrous and
ferric ions in water is studied in detail.
\end{abstract}
\maketitle

A wide variety of processes and reactions in electrochemistry, molecular 
electronics, and biochemistry have a common denominator: they involve a diabatic
electron transfer process from a donor to an acceptor \cite{Kuznetsov}. 
These reactions cover processes and applications as diverse as solar-energy conversion in 
the early steps of photosynthesis, oxidation-reduction reactions between a metallic electrode and solvated ions, 
and the I-V characteristics of molecular-electronics devices \cite{Ratner}. The key quantities of interest are  
the reaction rates (or, equivalently, the conductance) and the reaction pathways. 
Reaction rates, in the general scenario of Marcus theory \cite{Marcus,Newton,Warshel1}, have a thermodynamic 
contribution (the classical Franck-Condon factor, broadly related  
to the free energy cost of a nuclear fluctuation that makes the donor and 
the acceptor levels degenerate in energy), and an electronic-structure,  
tunneling contribution (the Landau-Zener term, related  
to the overlap of the initial and final states). 

We argue in the following that state-of-the-art first-principles
molecular dynamics calculations,
together with several algorithmic and 
conceptual advances, are able to describe with quantitative 
accuracy these diabatic processes, while
including the realistic description of the complex 
environment encountered, e.g. in nanoscale devices or at the interface between molecules 
and metals. 

Fig.~\ref{etrans} shows schematically an electron-transfer process and the free-energy diabatic surfaces
according to the picture that was pioneered by Marcus \cite{Marcus,Newton,Marcus_nobel,Chandler_book}.
In a polar solvent, 
the electron transfer process is mediated by thermal
fluctuations of the solvent molecules.
In the reactant state, the transferring electron is trapped at the donor site by solvent 
polarization; transfer might occur when the electron donor and acceptor 
sites become degenerate due to the thermal fluctuations of the solvent molecules.
To characterize the role of the solvent on the electron-transfer reaction, 
a reaction coordinate $\epsilon$ for a given ionic configuration is introduced, as
the energy difference between the product and reactant state at that configuration \cite{Warshel2}.
This definition of reaction coordinate captures the collective contributions from the solvent. 

\begin{figure}
\centerline{
\rotatebox{0}{\resizebox{2.8in}{!}{\includegraphics{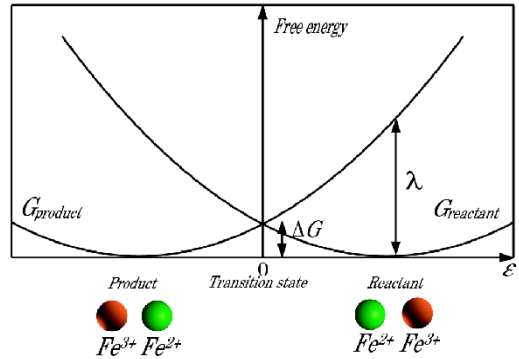}}}
}
\caption{Diabatic free-energy surfaces for ferrous-ferric electron transfer reactions. $\Delta$G is the
free energy barrier, and $\lambda$ is the reorganization energy. The reaction coordinate $\epsilon$
is the Marcus energy gap.}
\label{etrans}
\end{figure}

There have been a number of pioneering
classical molecular dynamic studies \cite{Warshel1,Warshel3,Kuharski} 
studying the reactions between aqueous metal ions. 
However, quantitative agreement has not been achieved with classical force fields.
The reorganization energy $\lambda$ (i.e. the free energy 
cost to reorganize the solvent molecules from the configurations at equilibrium with the product to the 
configurations at equilibrium with the reactant without electron transfer)
for the aqueous Fe$^{2+}$-Fe$^{3+}$ 
self-exchange reaction was found to be 3.6 eV for ions
5.5 \AA\ apart \cite{Kuharski}, while experimentally \cite{exp_reorg} (at 
the slightly shorter separation of 5.32 \AA) it is found to be 2.1 eV.  
Although there have been
studies of electron-transfer reactions including electronic polarization in classical
force-field potentials \cite{Warshel4,Warshel5}, full first-principles studies are required to 
describe realistically and quantitatively these reactions.  
Recently, an elegant grand-canonical density functional approach has been introduced to address this 
class of problems \cite{Sprik1,Sprik2}. This approach is, however, targeted at 
half-reactions for
a donor or an acceptor in contact with an electron reservoir. 

In this paper, we present
a novel technique to study electron-transfer reactions from first-principles molecular
dynamics, with ferrous-ferric self-exchange as a paradigmatic example. 
We use Car-Parrinello molecular dynamics \cite{CPMD,CPMD2} 
and spin-polarized DFT
in the PBE-GGA approximation \cite{Details}.
Fig.~\ref{procedure} shows schematically the sampling procedure  
used, following the lines of Ref.
\cite{Kuharski} for classical simulations.
An ionic trajectory is first generated with  
the ions in the (2+$r$) and (3-$r$) states of charge, respectively.
$r$ is an ``umbrella-sampling'' parameter used to explore different regions of the phase space. 
We then perform two separate runs with the electronic state constrained in the reactant or 
in the product configuration, and with the ions following
the afore-generated ionic trajectories. 
The reaction coordinate $\epsilon$ at every time step is thus given by 
the difference between the energies of the product and 
the reactant state. The probability distribution $P(\epsilon)$ is then calculated as:
\begin{equation}
P(\epsilon) = \frac{\sum\limits_{\tau} \delta_{\epsilon'(\tau),\epsilon} 
exp\{-\beta[E_{r}(\tau)-E_{s}(\tau)]\}}{\sum\limits_{\tau}exp
\{-\beta[E_{r}(\tau)-E_{s}(\tau)]\}},
\end{equation}
where $\epsilon'(\tau)=E_{p}(\tau)-E_{r}(\tau)$ is the reaction coordinate
at time $\tau$; $E_{r}(\tau)$, $E_{p}(\tau)$ 
and $E_{s}(\tau)$ are the energies of the system in the reactant, product and sampling oxidation states,
respectively, at time $\tau$, and $\delta_{\epsilon'(\tau),\epsilon}$ is the Kronecker delta.
The exponential term in the expression above restores the correct thermodynamical sampling according
to the energy surface $E_{r}$.
The free energy $G(\epsilon)$ is derived from the probability distribution 
$P(\epsilon)$ as $G(\epsilon)=-k_{B}$Tln$(P(\epsilon))$.

\begin{figure}
\centerline{
\rotatebox{0}{\resizebox{2.8in}{!}{\includegraphics{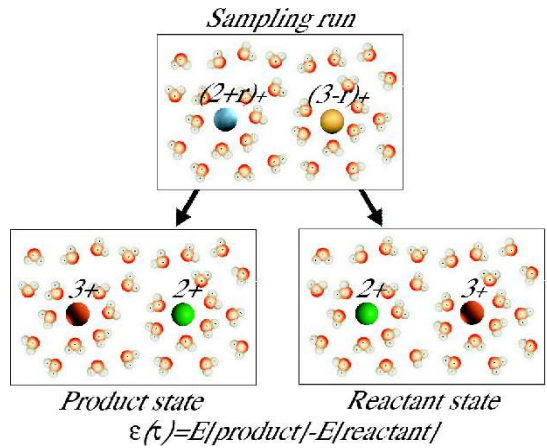}}}
}
\caption{Procedure used to calculate the diabatic free energy surfaces for electron transfer:
The reaction coordinate at each time step is calculated from the energy 
difference between the product and reactant states in the ionic configuration provided by the sampling run. 
The phase space is explored via the umbrella sampling parameter $r$, determining the oxidation state of
the ions.}
\label{procedure}
\end{figure}

It is of central importance to note that due to the lack of self-interaction correction in 
common exchange-correlation functionals, the transferring ($3d$ minority spin) electron will unphysically split
between the two ions. Moreover, 
to calculate the energy gap, we need to accurately calculate the total energy when the minority
spin electron localizes at either reactant or product site at any given ionic configuration.

In order to address these central problems, we first consider the simple case when oxidation states 
can be controlled trivially. This happens when two ions are infinitely apart; 
the two ions can be studied in separate simulation cells and the oxidation
states are controlled by simply changing the total number of electrons.
For this special case, we performed runs using
Fe$^{(2+r)+}$ and Fe$^{(3-r)+}$ (with r=0.0, 0.25,
0.50, 0.75 and 1.0), each solvated with 31 water molecules in the unit cell. 
We then carried out one Fe$^{2+}$ and one Fe$^{3+}$ 
run in the trajectory generated with Fe$^{(2+r)+}$,
and one Fe$^{2+}$ and one Fe$^{3+}$ run in the trajectory generated with Fe$^{(3-r)+}$. 
(When calculating energies 
of charged systems in periodic-boundary conditions, the Coulomb interaction of 
charges with their periodic images should be removed \cite{Makov}. In practice, these errors cancel 
out when calculating the energy gap, which is the difference in energy between systems with the same charge.)

Fig.~\ref{parabola} shows the resulting diabatic surfaces; 
the final result is obtained by integrating \cite{join}
\begin{equation}
F(\epsilon) = \frac{\sum\limits_{r} F_{r}(\epsilon) g_{r}(\epsilon)}{\sum\limits_{r} g_{r}(\epsilon)},
\end{equation}
where $F_{r}(\epsilon)$ is the slope of the free energy curve in the
different sections, each characterized by an umbrella-sampling parameter $r$, 
and the weighting factor 
$g_{r}(\epsilon)$ is $<\delta(\epsilon-\epsilon(\tau))>_{r}$.
Each simulation 
lasted 5 ps after accurate thermalization. For this special case, the trajectories
generated with Fe$^{(2+r)+}$ and Fe$^{(3-r)+}$ are independent; since each of them provides $n$ data
points, there will be $n^{2}$ energy gaps, providing high statistics and 
a very smooth free energy curve that 
fits accurately a parabola, with a
coefficient of determination ($R{^2}$) of 0.9996, and a reorganization energy of
1.77 eV.
Note that at the tail ends of each sampling region the statistical accuracy becomes lower -
this explains the slight deviations from a parabola seen in Fig.~\ref{parabola}.

\begin{figure}
\centerline{
\rotatebox{-90}{\resizebox{2.8in}{!}{\includegraphics{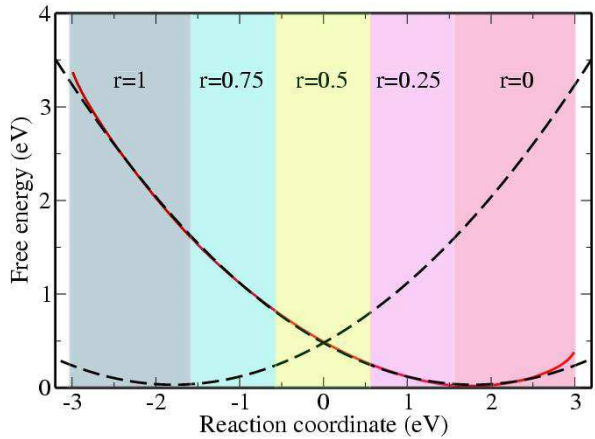}}}
}
\caption{Diabatic free energy surface for ferrous-ferric electron transfer in the special case when two ions 
are infinitely apart. The solid curve has been obtained from first-principles molecular dynamics. 
The dashed curves are mirror images, and correspond to a parabolic fit of the data.
Different shades indicate portions of the diabatic surface sampled with 
$r$=0, 0.25, 0.5, 0.75 and 1.}
\label{parabola}
\end{figure}

As mentioned earlier, the self-interaction errors of 
most exchange-correlation functionals result in a dramatic qualitative failure
in describing ions in different oxidation states when more than one ion is present. This failure can be
exemplified by the case of two iron ions in the 2+ and 3+ oxidation state.
When such a system is studied -- e.g. using PBE-GGA -- the HOMO electron will split between 
the two iron centers, to decrease its own self-interaction. This behavior takes place
irrespective of the chemical environment of the two iron centers; we observe it
for two isolated atoms, two hexa-aqua iron complexes, or two ions fully solvated.

We will show in the following that this failure can be corrected
by adding a penalty cost to ground states with non-integer occupation of the ion centers \cite{ldau}.
This same approach is also used to calculate the Marcus energy gap, where for
a given configuration we need to determine both the correct ground-state energy (with the
transferring electron in the reactant electronic configuration) and the first excited state
(with the transferring electron in the product electronic configuration).
We use and validate the following penalty functional
\begin{equation}
E[\{\psi_i\}] \rightarrow E [\{\psi_i\}]+ \sum\limits_{I}\frac{P^{I}}{\sigma_{I} \sqrt{2 \pi}}\int^{f_0^{I}-f_{\{\psi_i\}}^{I}}_{-\infty}  
\exp(-\frac{x^2}{2\sigma_{I}^{2}}) dx, 
\end{equation}
where $f_{\{\psi_i\}}^{I}$ is the largest eigenvalue of the minority-spin 
occupation matrix on ion I (calculated in this work by projecting the 
minority-spin Kohn-Sham orbitals on the $3d$ orbitals of an isolated iron atom), and $f_0^I$ is its target value.
To determine the optimal parameters, we separately calculated the minority-spin
occupation matrix for either a ferrous or ferric hexa-aqua ion embedded in a dielectric 
continuum ($\epsilon$=78) \cite{Damian}.
Then, we determine the parameters in the penalty functional so that the occupation matrices 
of the ferrous or ferric clusters are accurately reproduced once the two are studied in the same 
unit cell ($P^I$=0.54 eV, 
$f_0^I$=0.95 and $\sigma_I$=0.01 on the ferrous ion and $P^I$=-0.54 eV, $f_0^I$=0.28 and 
$\sigma_I$=0.01 on the ferric ion). 
We note that the target occupations for the minority-spin are not chosen exactly one or zero,
since the
orbital hybridization between the iron $3d$ orbitals and the lone pairs of the water molecules contributes to
the projection onto atomic orbitals (an alternative would be to use projections onto 
the $3d$ orbitals or the maximally-localized Wannier functions of a solvated iron, instead of an isolated atom).
When calculating the energy gap, the penalty functional contributions are 
taken away from the total energy; in any case, these effects are negligible since 
these contributions cancel out when calculating energy differences. 
Different constraints or penalties have been recently proposed for density-functional
calculations \cite{sit_aps05,mayeul,Troy,Scheffler}; we found our choice particularly robust,
but several variations on the theme can be envisioned.

A first, qualitative validation of this
penalty functional is performed examining the charge density obtained by subtracting from a calculation with
a ferrous and a ferric hexa-aqua ion in the same unit cell that of an isolated
ferrous hexa-aqua ion, and that of a ferric
hexa-aqua ion (a dielectric continuum surrounds the two clusters
to remove long-range electrostatic interactions between them).
When the penalty functional is applied, 
the charge density around the ions reorganizes itself so 
that it produces a charge density that is the exact superposition of that obtained from the 
two independent calculations.

We can make our validation quantitative by calculating the energy gap for the system described, using
two different penalty-functional calculations that impose to the HOMO electron to
localize first on one, then on the other ion.
This energy gap can also be calculated exactly with PBE-GGA
using the ``4-point'' approach \cite{4point}, provided that all long-range electrostatic interactions
are screened out.
The four calculations involve 
Fe$^{2+}$ in two Fe(H$_{2}$O)$_{6}$ geometries (A and B), and Fe$^{3+}$ in 
the same geometries; the energy gap is
$[E_{A}(Fe^{2+})+E_{B}(Fe^{3+})-E_{B}(Fe^{2+})-E_{A}(Fe^{3+})]$.
We choose one configuration in which the hexa-aqua ions are fully relaxed, and three carved out
from random steps in 
the molecular dynamics simulations. The energy gaps we obtained with the penalty
functional are 0.632, 0.569, 0.769 and 1.027 eV, 
in excellent agreement with the ``4-point'' values of 0.622, 0.542, 0.769 and 1.012 eV. 
It is worth mentioning that the energy gap is an excited-state property of the
system, and thus in principle outside the scope of density-functional theory, which is 
a ground-state theory. However, since the charge densities of the HOMO and LUMO do not overlap,
we can argue that all that is required is a description of the charge density that is locally correct
(the excited state has an electron locally in equilibrium around an iron,
oblivious of the other iron ion where it could sit more favorably). 

With these tools, we determined the diabatic free energy surfaces for two iron ions
separated by 5.5 \AA\, and solvated in 62 water molecules, in periodic boundary conditions. 
5.5 \AA\ was suggested \cite{Newton,optimal} to be the
optimal distance for electron transfer. 
We show our results in Fig.~\ref{parabola2}, together with a parabolic fit to the data. The reorganization
energy ($\lambda$) that we obtain is 2.0 eV \cite{finite}, in excellent agreement with the experimental value 
of 2.1 eV \cite{exp_reorg}.  The energy barrier $\Delta G\approx$0.49 eV,  
about a quarter of $\lambda$, as expected. We also note that since the structural
and electronic configurations of all microstates are available, reaction mechanism (e.g. inner vs. outer
sphere transfer) can be analyzed in detail by restricting the analysis to all configurations that
have a Marcus gap close to zero. Full first-principles accuracy also implies that bond-breaking and
bond-forming reactions can be followed in detail, or that protons can be explicitly introduced to study
the pH dependence of the reaction \cite{Rustad}.

\begin{figure}
\centerline{
\rotatebox{-90}{\resizebox{2.8in}{!}{\includegraphics{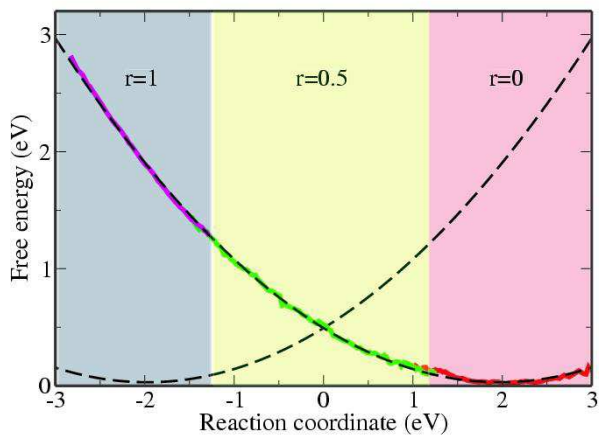}}}
}
\caption{Diabatic free energy surface for ferrous-ferric electron transfer when the two ions are 5.5 \AA\ apart.
Different color shades indicate portions of the diabatic surface sampled $r$=0, 0.5 and 1.
The right dashed curve is the parabolic fit of the data and the left dashed curve its
mirror image.}
\label{parabola2}
\end{figure}

In conclusion, we have demonstrated how it is possible to obtain Marcus diabatic surfaces from
first-principles molecular dynamics, where the entire system is treated quantum-mechanically,
with the accuracy and predictive power that this approach entails.
The case when two ions are at a finite distance requires special care in dealing with self-interaction
errors and excited-state energies. In response to these challenges, we developed
and validated a penalty functional that is able to control the oxidation states of ions, and
that describes accurately both the electronic ground state and the first
excited state where the electron is transferred to the other ion.
This approach can be successfully applied to a wide class of oxidation-reduction reactions, 
in solution (as it often happens in electrochemistry or biochemistry) or in the solid-state (intervalence charge-transfer).

We gratefully acknowledge support from
the Croucher Foundation (P.H.-L.S.), MURI grant DAAD 19-03-1-0169 (N.M.) and 
the MRSEC Program of the National Science 
Foundation under the award number DMR 02-13282 (P.H.-L.S.). 
Ref.~\cite{Chandler_book} was inspirational in addressing this problem from first-principles.
The calculations in this work have been performed using the Quantum-ESPRESSO package \cite{ESPRESSO}.
Computational facilities have been provided through NSF grant DMR-0414849 and PNNL grant EMSL-UP-9597.

\maketitle


\begin{thebibliography}{10}

\bibitem{Kuznetsov}
A. Kuznetsov and J. Ulstrup, {\it Electron Transfer Processes in Chemistry and Biology},
John Wiley and Sons, London, (1999). 

\bibitem{Ratner}
A. Nitzan and M. A. Ratner, Science {\bf 300}, 1384 (2003).

\bibitem{Marcus}
R. A. Marcus and N. Sutin, Biochem. Biophys. Acta {\bf 881}, 265 (1985).

\bibitem{Newton}
M. D. Newton and N. Sutin, Annu. Rev. Phys. Chem. {\bf 35}, 437 (1984). 

\bibitem{Warshel1}
A. Warshel and W. W. Parson, Annu. Rev. Phys. Chem. {\bf 42}, 279 (1991).

\bibitem{Marcus_nobel}
R. A. Marcus, Rev. Mod. Phys. {\bf 65}, 599 (1993).

\bibitem{Chandler_book}
D. Chandler, in {\it Classical and Quantum Dynamics in Condensed Phase Simulations},
edited by B.J. Berne, G. Ciccotti and D.F. Coker, pp. 1-66, World Sci, Singapore (1998).

\bibitem{Warshel2}
A. Warshel, J. Phys. Chem. {\bf 86}, 2218 (1982).

\bibitem{Warshel3}
A. Warshel, Nature, {\bf 260}, 679 (1976).

\bibitem{Kuharski}
R. A. Kuharski, J. S. Bader, D. Chandler, M. Sprik, M. L. Klein, and R. W. Impey, J. Chem. Phys.
{\bf 89}, 3248 (1988).

\bibitem{exp_reorg}
K. M. Rosso and J. R. Rustad, J. Phys. Chem. A {\bf 104}, 6718 (2000).

\bibitem{Warshel4}
A. Warshel and J. Hwang, J. Chem. Phys. {\bf 84}, 4938 (1986).

\bibitem{Warshel5}
G. King and A. Warshel, J. Chem. Phys. {\bf 93}, 8682 (1990).

\bibitem{Sprik1}
I. Tavernelli, R. Vuilleumier and M. Sprik, Phys. Rev. Lett. {\bf 88}, 213002 (2002).

\bibitem{Sprik2}
J. Blumberger, L. Bernasconi, I. Tavernelli, R. Vuilleumier and M. Sprik, J. Am. Chem. Soc. {\bf 126},
3928 (2004).

\bibitem{CPMD}
R. Car and M. Parrinello, Phys. Rev. Lett. {\bf 55}, 2471 (1985).

\bibitem{CPMD2}
K. Laasonen, A. Pasquarello, R. Car, C. Lee and D. Vanderbilt, Phys. Rev. B {\bf 47}, 10142 (1993).

\bibitem{Details}
O and H ultrasoft pseudopotentials are from the standard
PWscf distribution (H.pbe-rrkjus.UPF and O.pbe-rrkjus.UPF).
We generated a 16-electron iron ultrasoft pseudopotential (Fe.pbe-sp-van\_mit.UPF) 
using the Vanderbilt code (http://www.physics.rutgers.edu/~dhv/uspp/index.html). 
The wavefunctions and charge density cutoffs are 25 and 200 Ryd, respectively.
The deuterium mass was used in place of hydrogen mass to allow for
a larger time step of integration.
The fictitious mass ($\mu$) and the time step are 450 a.u. and
5 a.u., respectively. 
A Nose-Hoover thermostat was used on the both electrons and ions. The ionic temperature was set at 400 K.
See Ref. \cite{Grossman,sit}

\bibitem{Grossman}
J. Grossman, E. Schwegler, E. Draeger, F. Gygi and G. Galli, J. Chem. Phys. {\bf 120}, 300 (2004).

\bibitem{sit}
P. H.-L. Sit and N. Marzari, J. Chem. Phys. {\bf 122}, 204510 (2005).

\bibitem{Makov}
G. Makov and M. C. Payne, Phys. Rev. B {\bf 51}, 4014 (1995).

\bibitem{join}
J. S. Bader and D. Chandler, J. Phys. Chem. {\bf 96} 6423 (1992).

\bibitem{ldau} 
We note that GGA+U formulation would require a 
U of $\sim$ 11 eV to recover
localization of charge, but this large U would incorrectly affect the hybridization of
the Fe with the nearby water molecules.

\bibitem{Damian}
D. A. Scherlis, J.-L. Fattebert, F. Gygi, M. Cococcioni and N. Marzari, J. Chem. Phys. {\bf 124}, 074103 (2006).

\bibitem{sit_aps05}
P. H.-L. Sit and N. Marzari, Bull. Am. Phys. Soc. p. 1289 (2005).

\bibitem{mayeul}
M. d'Avezac, M. Calandra and F. Mauri, Phys. Rev. B {\bf 71}, 205210 (2005)

\bibitem{Troy}
Q. Wu and T. Van Voorhis, Phys. Rev. A {\bf 72}, 024502 (2005).

\bibitem{Scheffler}
J. Behler, B. Delley, S. Lorentz, K. Reuter and M. Scheffler, Phys. Rev. Lett. {\bf 94}, 036104 (2005);
J. Behler, B. Delley, K. Reuter and M. Scheffler, cond-mat/0605292.

\bibitem{4point}
S. F. Nelsen, S. C. Blackstock and Y. Kim, J. Am. Chem. Soc. {\bf 109}, 677 (1987).

\bibitem{optimal}
B. L. Tembe, H. L. Friedman and M. D. Newton, J. Chem. Phys. {\bf 76}, 1490 (1982).

\bibitem{finite} The electrostatic interaction with the periodic images for this unit cell
is less than 0.10 eV, assuming very conservatively that only the electrons screen the iron centers. In
reality ionic screening of the periodic images dominates even in the product electronic state,
leading to negligible electrostatic errors. Also, the pair correlation function of a ferrous or ferric ion
in water flattens around 5.5-6 \AA\, and the periodic images of each iron are 10.8 \AA\ apart.

\bibitem{Rustad}
J. R. Rustad, K. M. Rosso and A. R. Felmy, J. Chem. Phys. {\bf 120}, 7607 (2004).

\bibitem{ESPRESSO} S.  Baroni et al.,
http://www.quantum-espresso.org. 
                                                                                                                      
\end{thebibliography}
\end{document}